# Casimir Thrust Force on a Rotating Chiral Particle

Jianhui Yu,[1, 2, 3] Wenjin Long,[2, 3] Yu Zhang,[2, 3] Yongliang Li,[2, 3] Songqing Yang,[2] Wenguo Zhu,[2, 3, *] Huadan Zheng,[3, †] Yi Xiao,[2, 3] Jieyuan Tang,[2, 3] Heyuan Guan,[2, 3] Jiangli Dong,[2, 3] Huihui Lu,[2, 3] Wentao Qiu,[2, 3] Jun Zhang,[2, 3] Frank Tittel,[4] and Zhe Chen[1]

[1]*Guangdong Provincial Key Laboratory of Optical Fiber Sensing and Communications, Jinan University, GuangZhou, 520632, China*

[2]*Department of Optoelectronic Engineering, Jinan University, GuangZhou, 520632, China*

[3]*Key Laboratory of Optoelectronic Information and Sensing Technologies of Guangdong Higher Education Institutes, Jinan University, GuangZhou, 520632, China*

[4]*Department of Electrical and Computer Engineering, Rice University, Houston, Texas 77005, USA*

(Dated: October 5, 2018)

## Abstract

In the work, the thermal and vacuum fluctuation is predicted capable of generating a Casimir thrust force on a rotating chiral particle, which will push or pull the particle along the rotation axis. The Casimir thrust force comes from two origins: i) the rotation-induced symmetry-breaking in the vacuum and thermal fluctuation and ii) the chiral cross-coupling between electric and magnetic fields and dipoles, which can convert the vacuum spin angular momentum (SAM) to the vacuum force. Using the fluctuation dissipation theorem (FDT), we derive the analytical expressions for the vacuum thrust force in dipolar approximation and the dependences of the force on rotation frequency, temperature and material optical properties are investigated. The work reveals a new mechanism to generate a vacuum force, which opens a new way to exploit zero-point energy of vacuum.

*Introduction*-Vacuum quantum fluctuation gives rise to an attractive force between two parallel conducting plates, the so-called Casimir force [1], because the plates can modify the zero-point energy of the electromagnetic (EM) field. As a macroscopic quantum phenomenon, Casimir force has attracted widespread attentions and been intensively investigated [2–5] because it impinges on various fields ranging from fundamental physics to micromechanics, chemistry and biology when considering the micro-interaction of micro-/nano-objects, molecules and atoms [3, 6, 7]. The interaction between the object and the vacuum depends extremely on geometry and material property of the object, which opens a way of tailing Casimir force [8, 9]. For example, an ideal metal spherical shell [10], rectangular box [11], a metal needle and a metal plate with a hole [12], two dielectrics immersed in a medium of special optical properties [13], a chiral metamaterial [14], can result in repulsive force. The sign of Casimir force can be tuned by a separation between two plates with interleaved metal brackets [15] and the sign of Chern number in Chern insulator which constitutes the two plates [16]. Breaking translation symmetry with a corrugated surface on the object [17, 18] and breaking the reversal-time symmetry with a rotating particle near a surface [19] can give rise to lateral Casimir force. Restoring Casimir torque can also be induced by breaking azimuthal symmetry with anisotropic materials [20, 21].

Recently, dynamical interaction of an object with the thermal and vacuum fluctuation was investigated. Vacuum friction was theoretically predicted to exert on a surface and a small particle moving parallel to a surface [22–24], while rotational vacuum friction is on a rotating particle [25–27]. Lateral Casimir force can arise on a rotating particles near a surface because of the spin-orbit interaction [28, 29]. As well known, because of the chirality of the screw-blades, a rotating propeller can generate thrust force in water/air through collision between the rotating screw-blades and water/air molecules. In this letter, analogous to the propeller, we predict that a vacuum thrust force can be generated on a rotating chiral particle along the rotation axis through the interaction of the rotating chiral particles with thermal and vacuum fluctuations. Such vacuum thrust force originates from the rotation-induced symmetry-breaking and chirality-induced cross-coupling between electric and magnetic fields and dipoles. Using fluctuation dissipation theorem (FDT) [25, 26], we semiclassically derive the analytical expressions for the vacuum thrust force exerting on a rotating chiral particle in dipolar approximation. Numerically, we also investigate dependence of sign and magnitude of the force on rotation frequency, chiral coefficient and resonant frequency of chiral material.

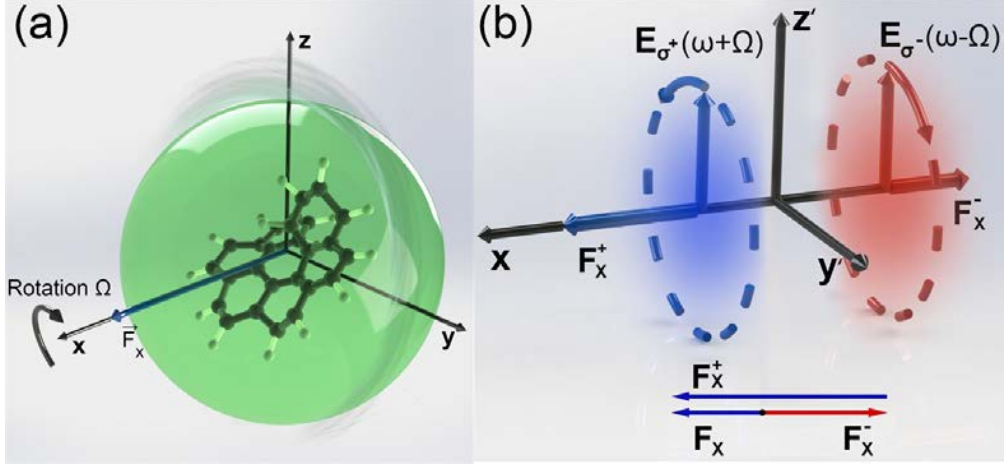

FIG. 1. (a) Schematic of vacuum thrust force on a rotating chiral particle at Ω in the lab frame. (b) illustration of the two origins of the vacuum thrust force in rotation frame: 1) rotation-induced frequency splitting, whereby the frequency of the circular polarized EM field of $\sigma^+$ increases to ω+Ω (blue), while the EM field of $\sigma^-$ reduces to ω−Ω (red), and 2) symmetry-breaking induced by the dispersion of polarizabilities and chiral parameter, which results in imbalance of two opposite spin-induced optical forces $\mathbf{F}^+$ and $\mathbf{F}^-$.

*Theoretical Description-* We consider a chiral spherical particle in vacuum, which is isotropic and small enough to be accurately described in dipole approximation through frequency-dependent chiral parameter $\chi(\omega)$, electric and magnetic polarizabilities $\alpha_e(\omega)$ and $\alpha_m(\omega)$. The electric and magnetic dipole moments of the nonrotating chiral particle can be expressed as $\mathbf{p}(\omega) = \alpha_e(\omega)\mathbf{E}(\omega) + i\chi(\omega)\mathbf{H}(\omega)$ and $\mathbf{m}(\omega) = -i\chi(\omega)\mathbf{E}(\omega) + \alpha_m(\omega)\mathbf{H}(\omega)$ [30]. According to the optical force on the chiral dipole [30], the two components of optical force $\mathbf{F}_{L_e} = \omega\gamma_e \langle\mathbf{L}_e\rangle$ and $\mathbf{F}_{L_m} = \omega\gamma_m \langle\mathbf{L}_m\rangle$ are, respectively, caused by electric and magnetic spins. Here, $\langle\ \rangle$ denotes average over time,
$$\gamma_e = -2\omega Im[\chi(\omega)] + \frac{\omega^4}{3c^3\epsilon_0} Re[a^e(\omega)\chi^*(\omega)]$$
,
$\gamma_m = -2\omega Im[\chi(\omega)] + \frac{\omega^4}{3c^3\mu_0} Re[a^m(\omega)\chi^*(\omega)]$ and $\mathbf{L}_{e(m)}$ is the electric(magnetic) spin density of the electromagnetic (EM) field. Interestingly, through the cross-coupling of the electric and magnetic fields by the chiral parameter $\chi$, the spin-induced optical force can provide a way of transferring spin angular momentum (SAM) of an external EM field to linear momentum of the particles along the direction of the spin of EM field. Additionally, note that the direction of the optical force depends only on the helicity of the circular polarized (CP) EM field $\hat{\sigma}^\pm = 1/\sqrt{2}(\hat{\mathbf{y}} \pm i\hat{\mathbf{z}})$ instead of the Poynting vector of the CP field. In the quantum frame of EM field, the vacuum SAM of the two opposite helicity $\hat{\sigma}^\pm$ of the CP field can be expressed as $\langle\mathbf{L}^\pm\rangle = \langle\mathbf{L}_e^\pm\rangle + \langle\mathbf{L}_m^\pm\rangle =$ $\pm\hbar\rho_0(\omega)N(\omega,T_0)\hat{\mathbf{x}}$, where Plank constant $\hbar$ denotes SAM carried by a single photon [31], the free-space photonic local density of states $\rho_0(\omega) = \omega^2/(\pi^2 c^3)$ [32] and the averaged photon number $N_T(\omega) = [e^{\hbar\omega/k_B T} - 1]^{-1} + 1/2$ in thermal vacuum at temperature $T_0$ [32]. Therefore, by integrating over frequency domain, the vacuum force induced by electric and magnetic spin of the two opposite helicity can be estimated as

$$\mathbf{F}^\pm = \mathbf{F}_{Le}^\pm + \mathbf{F}_{Lm}^\pm \propto \mp \int_0^\infty d\omega \left[ \begin{array}{c} \frac{\hbar\omega^4}{\pi^2 c^3} Im[\chi(\omega)] \\ -\frac{\hbar\omega^7}{6\pi^2 c^6} Re[\Upsilon(\omega)\chi^*(\omega)] \end{array} \right] N_{T_0}(\omega) \qquad (1)$$

Here, $\Upsilon(\omega) = \alpha_e(\omega)/\epsilon_0 + \alpha_m(\omega)/\mu_0$, $\mp$ correspond to the opposite helicity $\hat{\sigma}^\pm$, which indicates that the vacuum force induced by the CP of $\sigma^\pm$ the helicity has opposite direction. Therefore, for a nonrotating chiral particle, the symmetry of the vacuum leads to the equivalence of the vacuum spin momentum of the two opposite helicity, and thus balance of the spin-induced vacuum force associated with the opposite helicity, which contributes nothing to the vacuum thrust force. However, for a chiral particle rotating around the x axis with a rotation frequency of $\Omega$ as in Fig. 1(a), the rotation breaks the symmetry and thus results in the vacuum thrust force. In particular, in the rotating frame with the particles, due to angular Doppler effect [33–35] as in Fig. 1(b), the frequency of the CP field of $\hat{\sigma}^+$ helicity is increased to $\omega^+ = \omega + \Omega$, while the frequency of the CP field of $\hat{\sigma}^-$ helicity is decreased to $\omega^- = \omega - \Omega$ [28]. Accordingly, the electric polarizability $\alpha_e(\omega)$ of the nonrotating chiral particle will be split into two components, $\alpha_e(\omega^+)$ and $\alpha_e(\omega^-)$ for the CP fields of the opposite helicity. Similarly, the magnetic polarizability $\alpha_m(\omega)$ and chiral parameter $\chi(\omega)$ are, respectively, split into $\alpha_m(\omega^\pm)$ and $\chi(\omega^\pm)$. As a result, the splitting makes an imbalance in spin-induced vacuum force of the opposite helicity as in Fig. 1 (b), and thus leads to the vacuum thrust force.

To obtain an analytical expression for the vacuum thrust force more rigorously, we firstly consider an optical force exerting on a chiral particle in the dipole approximation, which can be written as [30, 36–38] $\mathbf{F} = \mathbf{F}_{edip} + \mathbf{F}_{mdip} + \mathbf{F}_{int}$, with an electric dipole force $\mathbf{F}_{edip} = \sum_i p_i \nabla E_i$, a magnetic dipole force $\mathbf{F}_{mdip} = \sum_i m_i \nabla H_i$ ($i = x, y, z$) and an interaction term $\mathbf{F}_{int} = -\frac{\omega^4}{6\pi c^3} \mathbf{p} \times \mathbf{m}$. The vacuum and thermal fluctuations causing the vacuum thrust force come from the two types: (i) fluctuation of dipole moments of the particles $\mathbf{p}^{fl}$ and $\mathbf{m}^{fl}$, (ii) vacuum fluctuation of the fields $\mathbf{E}^{fl}$ and $\mathbf{H}^{fl}$. As these two types of fluctuation are independent and uncorrelated,

$$F_{edip,x} = \sum_i \left\langle p_i^{fl} \partial_x E_i^{ind} + p_i^{ind} \partial_x E_i^{fl} \right\rangle \quad , \quad F_{mdip,x} = \sum_i \left\langle m_i^{fl} \partial_x H_i^{ind} + m_i^{ind} \partial_x H_i^{fl} \right\rangle \quad \text{and}$$

$$F_{int,x} = -\frac{\omega^2}{6\pi c^3} \sum_{ij} \epsilon_{xij} \left\langle p_i^{fl} m_i^{fl} + p_i^{pfl} m_j^{pind} + p_i^{mind} m_j^{fl} + p_i^{Efl} m_j^{Efl} + p_i^{Hfl} m_j^{Hfl} \right\rangle.$$ The first five parts of $F_{int,x}$ are, respectively, denoted by $F_{int,x}^{pmfl}$, $F_{int,x}^{pfl}$, $F_{int,x}^{mfl}$, $F_{int,x}^{Efl}$ and $F_{int,x}^{Hfl}$. Expressing the electric and magnetic dipoles induced by the field fluctuation as

$$\mathbf{p}^{ind}(\omega) = \overline{\overline{a}}_e^{eff}(\omega) \mathbf{E}^{fl}(\omega) + i \overline{\overline{\chi}}^{eff}(\omega) \mathbf{H}^{fl}(\omega) \tag{2}$$

$$\mathbf{m}^{ind}(\omega) = -i \overline{\overline{\chi}}^{eff}(\omega) \mathbf{E}^{fl}(\omega) + \overline{\overline{a}}_m^{eff}(\omega) \mathbf{H}^{fl}(\omega) \tag{3}$$

and the EM field induced by the dipole fluctuation as

$$\mathbf{E}^{ind}(\omega) = \frac{\omega^2}{c^2 \epsilon_0} \overline{\overline{G}}(\omega) \mathbf{p}^{fl}(\omega) + i\omega [\nabla \times \overline{\overline{G}}(\omega)] \mathbf{m}^{fl}(\omega) \tag{4}$$

$$\mathbf{H}^{ind}(\omega) = \frac{\omega^2}{c^2 \mu_0} \overline{\overline{G}}(\omega) \mathbf{m}^{fl}(\omega) - i\omega [\nabla \times \overline{\overline{G}}(\omega)] \mathbf{p}^{fl}(\omega) \tag{5}$$

, we obtain three components of the vacuum thrust force through FDT (see supplemental material [39] for details),

$$F_x^{dip+pmfl} = F_{edip,x} + F_{mdip,x} + F_{int,x}^{pmfl}$$
$$= \int_0^\infty d\omega \frac{\hbar \omega^4}{3\pi^2 c^3} Im \begin{bmatrix} \chi(\omega^+)(2N_{T_1}(\omega^+) + N_{T_0}(\omega)) \\ -\chi(\omega^-)(2N_{T_1}(\omega^-) + N_{T_0}(\omega)) \end{bmatrix} \tag{6}$$

$$F_{int,x}^{pfl+mfl} = F_{int,x}^{pfl} + F_{int,x}^{mfl}$$
$$= -\int_0^\infty d\omega \frac{\hbar \omega^7}{18\pi^3 c^6} \begin{bmatrix} Im[\Upsilon(\omega^+)] Im[\chi(\omega^+)] N_{T_1}(\omega^+) \\ -Im[\Upsilon(\omega^-)] Im[\chi(\omega^-)] N_{T_1}(\omega^-) \end{bmatrix} \tag{7}$$

$$F_{int,x}^{Efl+Hfl} = F_{int,x}^{Efl} + F_{int,x}^{Hfl}$$
$$= \int_0^\infty d\omega \frac{\hbar \omega^7}{18\pi^3 c^6} N_{T_0}(\omega) Re\left[ \Upsilon(\omega^+) \chi^*(\omega^+) - \Upsilon(\omega^-) \chi^*(\omega^-) \right] \tag{8}$$

where total vacuum thrust force $F_x^{tot} = F_x^{dip+pmfl} + F_x^{pfl+mfl} + F_x^{Efl+Hfl}$. Here, we can see that Eq. (6) derived from the FDT has the same form as the first term of Eq.(1) derived by the quantization of the vacuum EM field, while Eqs. (7) and (8) have the same form as the second term of Eq. (1). Additionally, from Eqs. (6-8), it can be seen more clearly that the vacuum thrust force originates from the rotation-induced splitting of polarizabilities and chiral parameter of the dipole.

*Numerical Results-* Since the optical chirality of natural material is small, we assume the rotating particle made of chiral metamaterial(CMM) [40]. For the CMM of the particle, the constitutive relation is defined as $\mathbf{D}(\omega) = \epsilon_0 \epsilon(\omega)\mathbf{E}(\omega) + i\frac{\kappa(\omega)}{c}\mathbf{H}(\omega)$ and $\mathbf{B}(\omega) = -i\frac{\kappa(\omega)}{c}\mathbf{E}(\omega) + \mu_0 \mu(\omega)\mathbf{H}(\omega)$ [14] where, for a $\Omega$ particle that is a canonical particle to model the CMM [14, 40], $\epsilon(\omega) = \epsilon_b + \frac{\Omega_e \omega_0^2}{\omega_0^2 - \omega^2 - i\gamma\omega_0}$ and $\mu(\omega) = \mu_b + \frac{\Omega_m \omega_0^2}{\omega_0^2 - \omega^2 - i\gamma\omega_0}$, and $\kappa(\omega) = \frac{\Omega_\kappa \omega_0 \omega}{\omega_0^2 - \omega^2 - i\gamma\omega_0}$ with the resonant strength coefficients $\Omega_e = 0.1560$, $\Omega_m = 0.0625$, $\Omega_\kappa = 0.0993$, the resonant frequency $\omega_0 = 1.8713 T Hz$ ($\lambda_0 = 1.01 mm$), damping rate $\gamma = 0.05463\omega_0$, background dielectric constant $\epsilon_b = 3.1736$ and permeability $\mu_b = 0.9798$[41]. Despite of the formula for a $\Omega$ particle, the dispersion formulas can be applied to the natural chiral material since the formulas of permittivity and permeability are in Lorentz shapes and the formula of chiral parameter is consistent with Condon model [42]. As shown in Figs. 2(a)-(c), the dipolar electric and magnetic polarizabilities $\alpha_e(\omega)$, $\alpha_m(\omega)$ and chiral parameter $\chi(\omega)$ of the chiral particle can be obtained by Mie scattering coefficients [38, 39]. Fig. 2(a)-(c) show the real and imaginary parts of the polarizabilities and chiral parameter with particle radius $R = 10\mu m$ (solid curve) and $R = 50\mu m$ (dashed curve), where the retardation effect leads to the resonant broadening in the $R = 50\mu m$ particle with respect to the $R = 10\mu m$ particle. To gain insight into the vacuum thrust force, we plot the integrand of the three components in Eqs. (6-8) and the total integrand of these components for the $R = 50\mu m$ particle as in Figs. 2(d)-(f), assuming a rotation frequency $\Omega/2\pi = 10 kHz$, temperature $T_0 = T_1 = 300K$. The inset in Fig. 2 (f) illustrates the rotating $\Omega$ particle and the vacuum force. In Figs. 2(d)-(f), we can see that the peak(dip) locates near resonant frequency $\omega_0$ due to the strong dispersion of the polarizabilities and chiral parameter near resonant frequency(indicated by dashed line), which dominates the vacuum thrust force. Therefore, a CMM having many resonant frequencies, which can be modeled as sum of many $\Omega$ particles [40], can enhance the vacuum thrust force.

The integrands of the three components in Eqs. (6-8) are proportional to the difference $A_j(\omega^+) - A_j(\omega^-)$. Here, $A_1(\omega) = Im[\chi(\omega)]N_T(\omega)$ for Eq. (6), $A_2(\omega) = Im[\Upsilon(\omega)]Im[\chi(\omega)]N_T(\omega)$ for Eq. (7), $A_3(\omega) = Re[\Upsilon(\omega)\chi^*(\omega)]$ for Eq. (8). Under the realistic condition of rotation frequency $\Omega \ll \omega_0$, $A_j(\omega^+) - A_j(\omega^-) \approx [2dA_j(\omega)/d\omega]\Omega$. Therefore, the integral of Eqs. (6-8) leads to the fact that $F_{int,x}^{pfl+mfl}$, $F_{int,x}^{Efl+Hfl}$ and $F_x^{tot}$ are proportional to rotation frequency $\Omega$ as in Fig. 3 (a). In Fig. 3, it is clear that the total vacuum force $F_x^{tot} \approx F_x^{dip+pmfl}$ since $F_x^{dip+pmfl}$ is 60~100 times larger than the components $F_{int,x}^{Efl+Hfl}$ and $F_{int,x}^{pfl+mfl}$. It can be also seen that $F_x^{dip+pmfl}$ is slightly larger in magnitude than $F_x^{tot}$ since $F_{int,x}^{Efl+Hfl}$ has opposite sign to $F_{int,x}^{Efl+Hfl}$. Similarly, the difference of vacuum photon number $N_T(\omega^+) - N_T(\omega^-)$ $\approx [2dN_T(\omega)/dT]\Omega T$, which leads to the linear temperature dependence of the vacuum force at $T \gg 0K$ as in Fig. 3(b). Note that, as shown in Fig. 3(c), this linear dependence is broken in low temperature range of 0 ~ 2K (indicated by red dashed line) and the total vacuum force approaches to constant value at zero temperature, $F_x^{tot}|_{T=0K} = -1.29 \times 10^{-28} N$. This is due to $N_T(\omega) \to 1/2$ when temperature $T \to 0$.

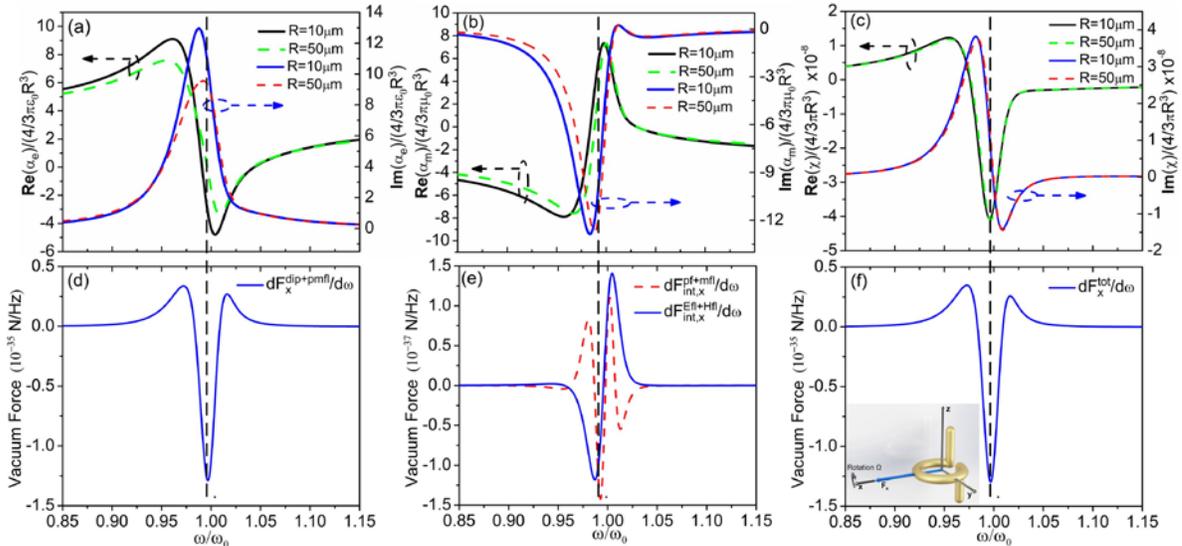

FIG. 2. (a)-(c) Real (left scale) and imaginary (right scale) parts of electric and magnetic polarizabilities $\alpha_e(\omega)$ in (a), $\alpha_m(\omega)$ in (b) and chiral parameter $\chi(\omega)$ in (c) with particle radius $R = 10\mu m$ (solid line) and $R = 50\mu m$ (dashed line; (d)-(f) Spectral density of the three force components in Eqs. (6-8) and total vacuum thrust force for the $R = 50\mu m$ particle at rotation frequency $\Omega = 10kHz$. The force spectral density are, respectively, $dF_x^{dip+pmfl}/d\omega$ in (d), $dF_{int,x}^{pfl+mfl}/d\omega$ (left scale) and $dF_{int,x}^{Efl+Hfl}/d\omega$ (right scale) in (c) and $dF_x^{tot}/d\omega = dF_x^{dip+pmfl}/d\omega + dF_{int,x}^{pfl+mfl}/d\omega + dF_{int,x}^{Efl+Hfl}/d\omega$ in (f). Inset shows the rotating $\Omega$ particle and the vacuum thrust force.

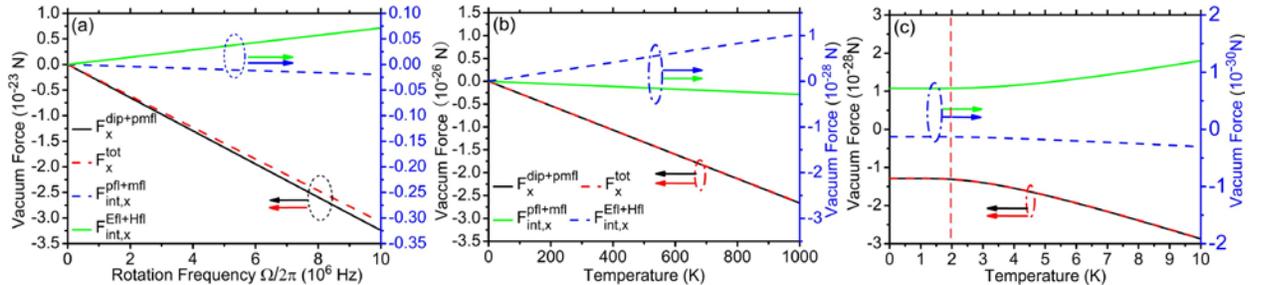

FIG. 3. Dependence of the vacuum thrust force, $F_x^{tot}$ (dashed red line), and its three components, $F_x^{dip+pmfl}$

(black solid line), $F_x^{pfl+mfl}$ (blue dashed line), $F_x^{Efl+Hfl}$ (green solid line), on rotation frequency $\Omega$ in (a) and temperature in (b) with its zoom-in at low temperature in (c). The dependence is plotted using the same parameters as in Fig.2, $R = 50\mu m$ and $\Omega/2\pi = 10\ kHz$.

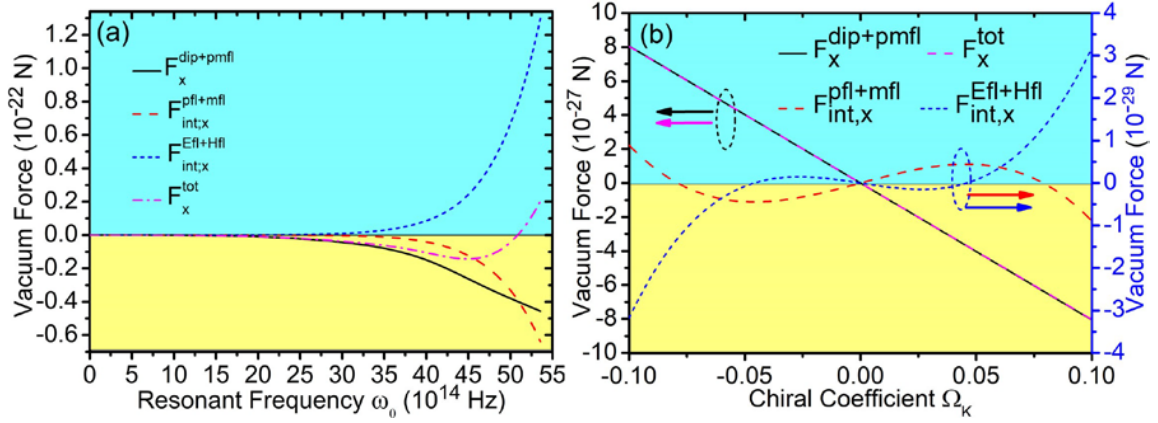

FIG. 4. (a) Dependence of the vacuum thrust force and its three components on resonant frequency $\omega_0$ of the chiral material with R=50nm and other parameters same as in Fig. 3;(b) their dependence on the chiral coefficient $\Omega_\kappa$ of the chiral material.

Interestingly, unlike natural materials, the CMM is a kind of metamaterial whose optical parameters can be tuned artificially by its structure [14, 40, 41]. Assuming the other material parameters unchanged, we investigate the sign and magnitude dependence of the vacuum thrust force on the resonant frequency $\omega_0$ and the chiral coefficient $\Omega_\kappa$, as in Figs. 4(a-b). In Fig. 4(a), with $\omega_0$ increase from $0.1T\ Hz$ to $856T\ Hz$ ($\lambda$: 18.8 $mm$ ~ 350$nm$), the magnitude of the components $F_{int,x}^{pfl+mfl}$ and $F_{int,x}^{Efl+Hfl}$ increase more rapidly than that of $F_{int,x}^{dip+pmfl}$ since $dF_{int,x}^{pfl+mfl}/d\omega$ and $dF_{int,x}^{Efl+Hfl}/d\omega$ are proportional to $\omega^7$, while $dF_{int,x}^{dip+pmfl}/d\omega \propto \omega^4$. It can be seen that the total vacuum thrust force $F_x^{tot} \approx F_x^{dip+pmfl}$ for $\omega_0/2\pi \leq 557T\ Hz$ ($\lambda = 538nm$) and $F_x^{tot}$ deviates highly from $F_x^{dip+pmfl}$ for $\omega_0/2\pi \geq 589T\ Hz$ ($\lambda = 509nm$). This is due to the sign of $F_{int,x}^{Efl+Hfl}$ is opposite to $F_{int,x}^{pfl+mfl}$ and $F_{int,x}^{dip+pmfl}$. With $\omega_0$ increase, the vacuum thrust force $F_x^{tot}$ reaches the maximum at $\omega_0/2\pi = 715T\ Hz$ ($\lambda = 419nm$) and turns from negative sign to positive one with crossing the zero value at $\omega_0/2\pi = 809T\ Hz$ ($\lambda = 371nm$). In the above calculation, the radius of particle is chosen as 50 nm to ensure the validity of dipolar approximation. As in Fig. 4, we can shrink down the size of the $\Omega$ particle to increase the resonant frequency and thus the vacuum thrust force.

Figure 4(b) plots the dependence of the three components, $F_x^{dip+pmfl}$, $F_{int,x}^{pfl+mfl}$, $F_{int,x}^{Efl+Hfl}$ and total vacuum thrust force $F_x^{tot}$ with $R = 50\mu m$, $\omega_0 = 1.817T\ Hz$. Except for the oscillation dependence of the components $F_{int,x}^{pfl+mfl}$ and $F_{int,x}^{Efl+Hfl}$, both two components $F_x^{dip+pmfl}$ and $F_x^{tot}$ are proportional to $\Omega_\kappa$. It is clear that the particle of the opposite chirality will suffer the vacuum thrust force in opposite direction. This provides a new mechanism to sort the chiral particle in vacuum at same rotation speed.

*Conclusions-* A vacuum thrust force is firstly predicted on the a rotating chiral particle, which originates from two factors: i) chirality-induced cross-coupling between electric and magnetic fields; ii) rotation-induced frequency splitting and dispersion of polarizabilities and chiral parameter. We derive the analytical expression for such vacuum force. Based on $\Omega$ particle model, numerical investigation shows that the vacuum force linearly depends on rotation frequency, temperature and chiral coefficient. The spectral force near resonant frequency of the chiral material dominates the thrust vacuum force. Increase in resonant frequency of the chiral material helps to enhance the vacuum force for the resonant frequency $\omega_0 \leq 557 T H$. The thrust vacuum force can be enhanced near a surface with surface plasmon polariton [27]. The work opens a new way to exploit the vacuum zero-point energy, for example, to implement a vacuum propeller for a spaceship and to separate the chiral particle in vacuum.

This work is supported by NSFC (61675092, 61475066, 61771222, 61705086, 61705087, 61705089); NSF of Guangdong Province (2016A030313079, 2016A030310098, 2016A030311019); Science and technology projects of Guangdong Province (2017A010102006, 2016A010101017, 2016B010111003); Science and Technology Project of Guangzhou (201707010396, 201707010253, 201704030105); Joint fund of pre-research for equipment, Ministry of Education of China (6141A02022124); Fundamental Research Funds for the Central Universities of China (11618413). Frank K. Tittel acknowledges the support by the Welch Foundation Grant No. C0568.

* WenguoZhu@jnu.edu.cn

† zhenghuadan@jnu.edu.cn